\documentclass[conference]{IEEEtran}
\usepackage{amsmath,amssymb,amsfonts}
\usepackage{algorithmic}
\usepackage{graphicx}
\usepackage{textcomp}
\usepackage{xcolor}
\usepackage{biblatex}
\usepackage{listings}

\title{SlimTCP: It's fast, but not because it's slim}

\author{\IEEEauthorblockN{1\textsuperscript{st} Mihai-Drosi Câju}
\IEEEauthorblockA{mcaju95@gmail.com}
\and
\IEEEauthorblockN{2\textsuperscript{nd} Costin Raiciu}
\IEEEauthorblockA{\textit{Universitatea POLITEHNICA București}\\
costin.raiciu@cs.pub.ro}
}
\vspace{10 mm}
\begin{document}

\maketitle

\begin{abstract}
In this paper, the authors explore the possibility of improving the performance of TCP/IP stacks in the context of data-center networks. This paper will focus particularly on the claim that simplifying the code-base of the stack increases its performance.
\end{abstract}

\begin{IEEEkeywords}
TCP, DPDK, performance optimizations
\end{IEEEkeywords}

\section{Introduction}
Ultra Ethernet\cite{hoefler2025ultra} provides a reliable-in-order channel over Ethernet\cite{ethernet} through its Reliable Ordered Delivery (ROD) profile. While TCP is a protocol whose performance is sensitive to packet loss or reordering. Thus, it should be possible to optimize a TCP\cite{tcp} stack in order to fully benefit from such a deployment. Several other technologies also exist that can provide such a channel for TCP, such as PFC\cite{pfc} coupled with ECMP\cite{ecmp} and IPoIB\cite{ipoib}. It is also conceivable that ECMP coupled with an over-provisioned network, such as one making use of 1.6Tb/s Ethernet\cite{1tbethernet} would also be one such channel. Therefore this paper will focus on the scenario where Ultra Ethernet is implemented inside a NIC, similar to RoCEv2\cite{rocev2}. In this scenario, the optimizations that would benefit the performance of a TCP/IP\cite{ipv4} stack involve stripping no longer needed features from the stack's fast-path. Thus, we present SlimTCP, a TCP/IP stack geared towards data-center use that implements the bare-minimum required to operate TCP and UDP\cite{udp} connections inside a data-center. We will also discuss the performance gains associated with a mock-setup.

\section{Design}
Given the nature of the channel in use, several extensions and optimizations present in TCP, as used over the internet, are no longer required. Thus, everything that assumed packet loss or reordering has been removed. RTO has been kept as SlimTCP was originally developed to run on top of EQDS\cite{eqds}. EQDS, in turn, offered imperfect packet delivery. The following extensions were not implemented: SACK\cite{tcpsack}, PAWS, TCP Timestamps\cite{tcphp}, congestion control\cite{tcpcong}, Fast Re-transmit\cite{tcpcong}, re-order buffers, etc. The only TCP extensions present are window scaling and MSS(Maximum Segment Size). WS(Window Scaling) was implemented to enhance performance at high BDPs(Bandwidth Delay Products). While MSS was implemented in order to test the stack's performance at varying segment sizes.

\section{API}
The control path follows the POSIX socket API\cite{posix}, with equivalent functions for the socket life-cycle having alternative implementations. Several extra socket options were implemented for performance testing, these allowed directly setting the window size and transmitted and received maximum segment sizes. For clarity, the API functions are:
\begin{lstlisting}
int ndpip_socket(int domain, int type,
    int protocol);
    
int ndpip_bind(int sockfd,
    const struct sockaddr *addr,
    socklen_t addrlen);
    
int ndpip_listen(int sockfd, int backlog);

int ndpip_connect(int sockfd,
    const struct sockaddr *addr,
    socklen_t addrlen);
    
int ndpip_accept(int sockfd,
    struct sockaddr *addr,
    socklen_t *addrlen);
    
int ndpip_close(int sockfd);

int ndpip_setsockopt(int sockfd, int level,
    int optname, const void *optval,
    socklen_t optlen);
    
int ndpip_getsockopt(int sockfd, int level,
    int optname, const void *optval,
    socklen_t optlen);
\end{lstlisting}

It should be noted that SlimTCP is internally called ndpip, as such all function calls are prefixed with ndpip\_. The POSIX API is not replaced with methods such as system call interception or symbol redirection. Rather, the developer must voluntarily opt to use SlimTCP in their source-code.

To enhance the performance of the data-path, an alternative API was implemented. One disadvantage of the POSIX equivalent API is that it requires buffer copies between the NIC's DMA\cite{pcie} buffers and the application's. In contrast, by using the zero-copy alternative, the application writes data directly to the buffers that will be placed in the NIC's TX ring. While buffers in the NIC's RX ring are delivered directly to the application. For fairness of comparison with other stacks, the POSIX-like API was used in the benchmarks. The data-path API is as follows:
\begin{lstlisting}
int ndpip_recv(int sockfd,
    struct ndpip_pbuf **pb, uint16_t count);
    
int ndpip_send(int sockfd,
    struct ndpip_pbuf **pb, uint16_t count);
    
ssize_t ndpip_read(int sockfd,
    void *buf, size_t len);

ssize_t ndpip_write(int sockfd, void *buf,
    size_t len);
    
int ndpip_free(int sockfd,
    struct ndpip_pbuf **pb, size_t len);
    
size_t ndpip_alloc(int sockfd,
    struct ndpip_pbuf **pb, size_t len);
    
int ndpip_prepare(int sockfd,
    struct ndpip_pbuf *pb);
\end{lstlisting}

While using the zero-copy API, on the transmit path, an application must first allocate packets from the socket's packet buffer pool. These are then prepared for transmission. This involves setting metadata, adjusting the buffer size and setting some protocol fields. The prepare function is provided due to the possibility that the send function may reject sending the packet buffers. This, in turn, returns ownership of the buffers to the application, unaltered. The send function takes as input an array of packet buffers and either changes their ownership to the NIC's TX ring or returns them to the application if the send window is full. The send function also writes additional protocol fields that can only be known at send time and. If the socket's protocol is TCP, it inserts the packet in the re-transmit ring. While using, the POSIX equivalent API, a simple ndpip\_write suffices, with no additional set-up. The POSIX equivalent API uses the zero-copy API as its backend.

In order for an application to receive packets from a socket's receive ring, it must call the recv function. This, in turn, will transfer the ownership of the packet buffers to the application. The application must then consume and free them. To read them using the POSIX equivalent API a ndpip\_read as used on Unix is sufficient.

To asynchronously wait for events on a set of sockets an epoll-equivalent API is also provided.

\section{Architecture}
The stack functions entirely in user-space with DPDK\cite{dpdk} as its networking driver. This choice was made in order to be able to write a stack from scratch while avoiding the complexity of writing kernel code.

The stack's functionality is spread across two cores. The first is managed by the application and is responsible for the reading and writing of data segments as well as the socket life cycle and enqueueing packets to the NIC. The second core hosts the worker thread that processes all incoming packets according to each protocol's state machine as well as timers and protocol replies.

In order to properly assess the performance of TCP under the given hypothesis, technologies such as TSO\cite{tso}, LRO\cite{lro} and RSS\cite{rss} were not used as they would have artificially sped-up the stack's performance. Therefore, packet reception and transmission were exclusively single-core and the stack provided a one-to-one mapping between processed buffers and on-wire data. The stack can make use of transmit and receive checksum offloading with a fallback to software implementations.

In order to maximize performance while sending and receiving buffers, the passing of buffers between stack and application makes use of lock-less single-producer single-consumer ring buffers\cite{rings}. These ring buffers are in use on the re-transmission ring for TCP on the send path, while on the receive path, buffers are indexed in such rings until ready for consumption by the application.

To speed-up the hash-tables in use while providing some security guarantees, the hash functions in use are based on XXH32\cite{xxh32}.

The epoll\cite{epoll} implementation makes use of a busy wait API. This design choice was made due to its simplicity and the reduced invocation of the kernel for synchronization primitives on the receive path.

\section{Optimizations}
Due to TCP being flexible on the timing and distribution of ACKs, pure ACKs are sent only once per receive burst for each processed socket. This has a similar effect to LRO. The same delay is used for the freeing of the re-transmission ring once ACKs are received.

In order to make efficient use of locking, SlimTCP buffers received packets onto a per-socket array before being fed to the socket. This array is distinct from the one that pushes to the sockets receive ring. Given that each burst has a static maximum size, bounds checking on this array is not required and avoids extra memory reads. This optimization allows the stack to lock only once per socket per burst and avoid superfluous locks. This is the case only for TCP as UDP does not have state on receive that requires locking.

As a general rule, pointer dereferencing was kept to a minimum so as to optimise cache use. It was noticed that memory intensive instructions had a high performance penalty, especially on cache misses.

\section{Evaluation}
Evaluation was done on top of a bare-metal set-up with two hosts each with 10 cores and 32GiB of RAM, the CPU model was Intel(R) Xeon(R) CPU E5-2670 v2. The hosts were connected through a switch. The installed NICs were Broadcom StingRay PS225, these allowed for a total bandwidth of up to 25Gb/s and up to 68Mpps per port. In the testbed only one port was used as the packet rate was bound by the CPU rather than the NIC.

The evaluation metrics consisted of the achieved packet rate and goodput while varying the number of connections and transmitted segment sizes.

Three stacks were benchmarked, SlimTCP, F-Stack\cite{fstack} and mTCP\cite{mtcp}. F-Stack and mTCP were patched so as to not verify checksums on the receive path. Several other features were patched-out as well, such as LRO, TSO and RSS. Timestamps were removed so as to keep the segment size constant across stacks. mTCP was modified so that it's core affinity matched that of F-Stack and SlimTCP, namely one application core and one worker core. Another modification made to mTCP allowed setting the MSS through environment variables.

To test the three stacks, three separate benchmark programs were written corresponding to each stack. The programs functioned similarly to iperf\cite{iperf} while allowing to set the transmitted segment length. That is, they implemented a server that received TCP data from a raw socket by using the equivalent read system call of each stack. The server also implemented event polling on multiple sockets in the second experiment. The client would call the equivalent write system call in a loop with an uninitialized buffer. The server would then measure the resulting goodput.

All stacks were compiled with the same optimizations(-g -O3 -march=sandybridge) and compiler version(GCC 9\cite{gcc}). So as to maintain consistency and reproducibility, the Nix\cite{nix} package manager was used to compile and deploy the testbed. Each stack used its own version of DPDK due to it not being feasible to port each version to the same DPDK version.

Two benchmarks were performed. One by varying the segment size while using just one connection. And the other by keeping a fixed 1400 MSS and varying the number of connections.

Each transmitter receiver combination of stacks was tested.. In total 9 tuples of stacks and protocols were benchmarked.

\section{Results}

\begin{table*}
\vspace{0.1 in}
\centering
\begin{tabular}{ *{6}{c} }
\multicolumn{6}{c}{MEAN goodput(b/s) by segment size} \\
\hline
tx & rx & MSS 64B & MSS 256B & MSS 512B & MSS 1460B \\
SlimTCP & SlimTCP & 1.9G & 6.8G & 13.0G & 23.7G \\
f-stack & f-stack & 670.5M & 2.1G & 2.7G & 3.2G \\
mtcp & SlimTCP & 2.1G & 8.1G & 13.8G & 16.5G \\
SlimTCP & mtcp & 336.2K & 1.7M & 7.3G & 1.5M \\
mtcp & mtcp & 1.3G & 4.4G & 8.7G & 12.4G \\
SlimTCP & f-stack & 675.9M & 1.7G & 2.3G & 190.7K \\
f-stack & SlimTCP & 410.6M & 5.2G & 9.2G & 15.7G \\
mtcp & f-stack & 668.2M & 1.8G & 2.3G & 3.0G \\
f-stack & mtcp & 123.4M & 576.1M & 8.0G & 776.8M \\
\end{tabular}
\end{table*}

\begin{table*}
\centering
\begin{tabular}{ *{6}{c} }
\multicolumn{6}{c}{MEAN total goodput(b/s) by number of connections} \\
\hline
tx & rx & 1 con & 1024 con & 2048 con & 4096 con \\
f-stack & mtcp & 2.4G & 12.3G & 9.7G & 0.0 \\
f-stack & f-stack & 3.2G & 2.4G & 1.4G & 0.0 \\
mtcp & mtcp & 12.3G & 16.0G & 16.0G & 15.2G \\
SlimTCP & SlimTCP & 23.7G & 23.7G & 23.7G & 23.6G \\
f-stack & SlimTCP & 15.5G & 7.4G & 5.1G & 3.3G \\
mtcp & f-stack & 3.0G & 2.5G & 1.4G & 0.0 \\
mtcp & SlimTCP & 16.6G & 20.3G & 19.7G & 19.1G \\
SlimTCP & mtcp & 1.7M & 968.3M & 1.3G & 1.3G \\
SlimTCP & f-stack & 193.3K & 267.7M & 0.0 & 0.0 \\
\end{tabular}
\end{table*}

The most performant tuple on one connection for larger MSS's(above or equal to 1024) was SlimTCP as a sender coupled with SlimTCP as as receiver,. It was also the only tuple to saturate the link at 23.6Gb/s goodput. While mTCP coupled with SlimTCP was faster for smaller MSS's.

The most scalable tuple for the second benchmark was SlimTCP coupled with with SlimTCP. While SlimTCP as a sender coupled with other stacks would trigger RTO at times. This may have been due to the sender being too fast and the NIC's receive rings overflowing.

\section{Discussion}
One drawback as of this point in SlimTCP is its lack of receive window scaling, this allows it to function only when the application thread consumes the packet buffers for each socket fast enough not to trigger an assert on a full ring buffer or depleted memory pool.

Although SlimTCP shows promising performance on this testbed and out-competes f-stack and mTCP in most tested scenarios, it is still questionable if the hypothesis presented in this paper was validated by it. This is the case as it is unclear whether a combination of bugs, performance optimizations and non-conformity had a greater effect on the stack's performance and scalability rather than the optimization for the channel's hypothesis.

The performance drop in SlimTCP for a sender having segments below 1024 can be explained by the contention over the socket structure lock between the application thread that sends segments and the worker thread that handles ACKs. This design decision was taken in order to mimic Linux's behavior and distribute the workload in the case of bidirectional traffic. However, this turned out to degrade the performance of the sender with too frequent ACKs. A better approach would have been to mimic mTCP instead and move the sender on the worker thread as well. This would have removed the need of a lock between the sender and receiver on the fast-path. It is still questionable how this would have impacted performance in a real-world application.

Given that SlimTCP is still in it's infancy as a TCP/IP stack, some bugs may still be present. Thus, a fair comparison with a production-hardened stack may be unfair, thus F-Stack may still be better. While, even mTCP is more mature, although it is still a research-mostly stack.

Although one of the premises of a slimmed-down TCP was its compatibility with internet-attached devices without requiring middle-boxes. The lack of features such as Fast Retransmit, SACK and congestion control, would make it the weakest link in the connection chain. Thus, an alternative would be to make use of a middle-box that would proxy connections between the data-center and the internet.

Furthermore, although attempts were made to limit each stack's usage to one core, SlimTCP has its sender mostly on the application core. While the workload is the same for each stack, the lightweight nature of the workload may have given SlimTCP an unfair advantage. This remains to be evaluated by an application-oriented benchmark.

It should also be noted that at the time of writing, SlimTCP makes use of a, possibly, superfluous amount of RAM(2GiB). This was calculated in order to fit the maximum receive window for TCP(1GiB) for both the sender and receiver. Memory optimizations may be attempted in the future. 

The claim that a simpler stack would benefit performance seemed to have been supported by the limited resource of a CPU's L1i cache. As a counter-argument, it can be claimed that, although code paths may experience L1i evictions, they may still keep a large part of performance compared to L1d misses as the later frequently involves a RAM fetch while the former may involve just an L2 fetch. By code size measurements between mTCP and SlimTCP it was noticed that both had code-sizes around 1KiB for the receive path applied to each TCP segment. Even in the case that additional code-size reductions were performed they would not have brought any performance benefits as L1i on modern systems (in the order of 10's of KiB)\cite{amdl1} would have been under-used. Even if the entire stack nor the drivers may not fit within L1i, it is superfluous in holding a performance critical loop, as can be found on the receive path of both stacks.

The claim may have been even supported by the increased performance brought by reduced branching. However, modern CPUs employ complex branch-prediction to mitigate this effect. And thus, this argument fails as well.

Even under the most rigorous conditions a measurement of a stack's performance may be impossible to correlate with its code complexity. As together with code complexity there are always natural variances in implementation details. That is, code quality evolves together with code quantity in software development. While the two can not be separated and quality can not be separated from human or technical factors, it becomes impossible to clearly discern that reduced complexity would have resulted in an improved performance. More generally this question can be asked as: Given a certain decrease in code complexity, would a stack be possible that has a greater performance? Answering this question is close to impossible using an exhaustive approach as there are far to many other factors to take into account.

This stack, while inheriting the dependency on DPDK of F-Stack and mTCP, manages to compete with and outperform them in certain conditions. However, given the monolithic nature of modern operating system kernels its adoption is close to impossible.

\section{Conclusion}
The claim that a simpler protocol and or TCP/IP stack would improve performance remains ambiguous, at least under the premises of sane software engineering.

Although, a faster stack has been produced under laboratory and possibly future data-center conditions, its lack of feature parity with other stacks, lack of code maturity as well as difficulty in implementation make it of limited if any use.

However, a point has been made in this paper by reinforcing the common wisdom that quality work is a major contributor to software performance.

The source-code for the performance benchmark is at \url{https://github.com/CajuM/eqds-cloudlab} while the source-code for SlimTCP is at \url{https://github.com/CajuM/lib-ndpip}.

\section{Future Work}
A better validation of the stack's performance would consist of benchmarking real-world applications on top of it.

This validation will automatically include bidirectional traffic and allow for the opportunity to investigate weather moving parts of the sender to the receiver thread would improve performance.

Focus may also be placed on porting the stack to other applications such as in-hypervisor TCP accelerators or unikernel TCP stacks. A Rust\cite{rust} re-write may also be undertaken.

\printbibliography

\end{document}